\documentclass[letterpaper,10pt,conference]{ieeeconf}
\IEEEoverridecommandlockouts
\usepackage[utf8]{inputenc}
\usepackage{algorithm}
\usepackage{algpseudocode}
\usepackage[english]{babel}
\usepackage{color}
\usepackage{colortbl,tabulary,tabularx,multirow}
\usepackage{ltxtable}
\usepackage{comment,paralist}
\usepackage{amsmath}
\usepackage{amssymb}
\usepackage{stmaryrd}
\usepackage{braket}
\usepackage{textcomp}
\usepackage{listings}
\usepackage{theorem}
\usepackage{alltt}
\usepackage{graphicx}
\usepackage[colorlinks,linkcolor=blue,citecolor=blue,urlcolor=blue]{hyperref}
\usepackage{multicols}
\usepackage{balance}
\usepackage{amssymb}
\usepackage{graphics} 
\usepackage{epsfig} 

\everymath{\displaystyle}

\definecolor{cgreen}{rgb}{0,0.6,0}
\renewcommand{\baselinestretch}{1.0}

\renewcommand{\Re}{\mathbb{R}}
\renewcommand{\c}{\ensuremath{\operatorname{c}\!}}

\newcommand{\benum}{\begin{enumerate}}
\newcommand{\eenum}{\end{enumerate}}
\newcommand{\nc}{\newcommand}
\newcommand{\rnc}{\renewcommand}
\newcommand{\bvb}{\begin{verbatim}}
\newcommand{\evb}{\end{verbatim}}

\nc{\bq}{\textbf}
\nc{\m}{\textrm}
\nc{\bb}{\mathbb}

\nc{\til}{\texttildelow}
\nc{\be}{\begin{equation}}
\nc{\ee}{\end{equation}}
\nc{\dps}{\displaystyle}
\rnc{\l}{\left(}\rnc{\r}{\right)}
\nc{\lc}{\left\{}\nc{\rc}{\right\}}
\nc{\lb}{\left[}\nc{\rb}{\right]}
\nc{\ba}[1]{\begin{array}{#1}}
\nc{\ea}{\end{array}}
\nc{\ra}{\rightarrow}
\nc{\li}{\left |}
\nc{\ri}{\right |}
\nc{\pde}[2]{\frac{\partial #1}{\partial #2}}
\nc{\ode}[2]{\frac{d #1}{d #2}}
\nc{\odee}[3]{\frac{d^{#3} #1}{d #2^{#3}}}
\nc{\pdee}[3]{\frac{\partial^{#3} #1}{\partial #2^{#3}}}
\nc{\bn}{\begin{enumerate}}
\nc{\en}{\end{enumerate}}
\nc{\bt}{\begin{theorem}}
\nc{\et}{\end{theorem}}
\nc{\y}[1]{\lambda_{#1}}
\nc{\ninf}{{\oplus}^{-\infty}}
\nc{\pinf}{{\oplus}^{+\infty}}
\nc{\nninf}{{\otimes}^{-\infty}}
\nc{\ppinf}{{\otimes}^{+\infty}}
\nc{\ir}{\mathbb{I}\mathbb{R}}

\nc{\ep}{\mathcal{E}_{P}}
\nc{\mr}{\mathcal{M}_{r}}
\nc{\mfa}{\mathcal{M}_{f,a}}
\nc{\mfp}{\mathcal{M}_{f,p}}
\nc{\mt}{\m{T}}
\nc{\F}{\mathbb{F}}

\newtheorem{theorem}{Theorem}[section]
\newtheorem{lemma}[theorem]{Lemma}

\title{\LARGE \bf
Credible Autocoding of Fault Detection Observers
}

\author{Timothy E. Wang$^{1}$, Alireza Esna Ashari$^{2}$, Romain J. Jobredeaux$^{3}$, and Eric M. Feron$^{4}$
\thanks{*This article was prepared under support from NSF Grant CNS - 1135955 ``CPS: Medium:
Collaborative Research: Credible Autocoding and Verification of Embedded Software
(CrAVES)'', NASA Grant  NNX12AM52A ``Validation Elements For Loss-of-Control Recovery Operations (VELCRO)'', the Army Research Office under
MURI Award W911NF-11-1-0046, and ANR ASTRID project VORACE.}
\thanks{$^{1}$Timothy Wang is a PhD student in the Department of Aerospace Engineering, Georgia Tech, Atlanta, GA 30332, USA
        {\tt\small timothy.wang@gatech.edu}}%
\thanks{$^{2}$Alireza Esna Ashari is a Post-doctoral Researcher in the Department of Aerospace Engineering,
	Georgia Tech, Atlanta, GA 30332, USA
        {\tt\small aeae3@mail.gatech.edu}}%
\thanks{$^{3}$Romain Jobredeaux is a PhD student in the Department of Aerospace Engineering at Georgia Tech
        {\tt\small rjobredeaux3@gatech.edu}}%
\thanks{$^{4}$Eric Feron is the Dutton/Ducoffe Professor
	with the Department of Aerospace Engineering, Georgia Tech
        Atlanta, GA 30332, USA
        {\tt\small feron@gatech.edu}}%
}

\begin{document}

\maketitle
\thispagestyle{empty}
\pagestyle{empty}

\begin{abstract}
In this paper, we present a domain specific process
to assist the verification of observer-based fault detection software.
Observer-based fault detection systems, like control systems, yield invariant properties of quadratic types.
These quadratic invariants express both safety properties of the software such as the boundedness of
the states and correctness properties such as the absence of false alarms from the fault detector. We
seek to leverage these quadratic invariants, in an automated fashion, for the formal verification of
the fault detection software.  The approach, referred to as the credible autocoding framework~\cite{wjf13arxiv},
can be characterized as autocoding with proofs. The process starts with the fault detector model,
along with its safety and correctness properties, all expressed formally in a synchronous modeling
environment such as Simulink. The model is then transformed by a prototype credible autocoder into both
code and analyzable annotations for the code.  We demonstrate the credible autocoding process on a running example
of an output observer fault detector for a 3 degree-of-freedom (3DOF) helicopter control system.
\end{abstract}

\section{Introduction}
The safety of dynamic systems has attracted attention over years. Many studies on fault detection of safety-critical systems are reported in recent years \cite{hwang2010survey,gertler1988survey,frank1990fault,frank1997survey,isermann1997supervision,campbell2004auxiliary,esna2011auxiliary, niemann2000design,niemann1997robust,esna2010active}.
Most of those studies detection developed observer-based fault detection methods, which are suitable for online fault detection in the case of abrupt faults. The observer provides analytic redundancy for the dynamics of the system. Comparing the input-output data with the nominal data we obtain from the model of the system, we conclude whether or not the system is in nominal mode. Possible faults are modeled as additive inputs to the system. Such an additive fault changes the nominal relations between inputs and outputs.
A summary of the recent developments in this domain can be found in \cite{ding2008model,patton1999robust,isermann2005fault}.
Nowadays, observer-based fault detection methods are usually implemented as software on digital computers. However, there is usually a semantic gap between fault detection theory and
software implementation of those methods. Computation errors may cause incorrect results. Also, engineers with little or no background in control may need to test and modify the software. Thus there is a need to express fault detection semantics at the level of software. Additionally, such an endeavor can help verify systematically that the software works correctly based on the theory and the initial design.

In this paper, we present an automated process of applying
control-theoretic techniques towards the verification of observer-based fault detection software.
We extend our previous works~\cite{wjf11arxiv},~\cite{wjfdasc11},~\cite{heber},~\cite{wjf13arxiv} for controller systems
to fault detection systems.
The idea of using domain-specific knowledge in software verification is not new.
However, the application of
system and control theory to control software verification can be traced back
to relatively recent works like~\cite{feroncsm10} and~\cite{ferretesop04}.
In these papers, the authors presented a manual example of documenting a controller program with a
quadratic invariant set derived from a stability analysis of the state-space representation of the
controller.
Since then, we have progressed towards creating
an automated framework that can rapidly obtain and transform high-level functional properties of the control system into logic statements that are embedded into the generated code in the form of comments.
The usefulness of these comments comes from their potential usage in the automatic verification of the code.
We will refer to the logic statements as ``annotations''
and the generated code with those comments as ``annotated code''.
We named the framework \emph{credible autocoding}
as it is a process to rapidly generate the software as well as
the annotations that guarantee some functional properties of the system.
The realization of the framework is a prototype tool that we have built and applied to control systems such as
a controller for the 3 degrees of freedom helicopter Quanser~\cite{Quanser}.
For this paper, we have further refined the prototype to handle the addition of a
fault detection system running along with the Quanser controller.

The paper is organized as follows:
first we introduce credible autocoding framework and the details of its implementation;
this is followed by a mathematical description of the fault detection method used in our running example;
finally, we describe the credible autocoding process that has been
further extended for fault detection systems and the automatic verification of the annotated code produced by the
credible autocoder.

\section{Credible Autocoding}

Credible autocoding is an automatic or semi-automatic process
that transforms a system that is initially expressed in a
language of high-level of abstraction, along with the
mathematical proofs of its good behavior, into code, annotated with said mathematical proof.
The initial level of abstraction could be a
differential equation of the system and
the final level could be the software binary.
For the prototype implementation of our framework, we picked Simulink as
the starting point and C code as the final output.
Regardless of the input and output languages, the main contribution from this prototype
is the automatic translation of Lyapunov type stability proofs into axiomatic semantics
for the output code.

Axiomatic semantics is an approach to reason about the correctness of program
that traces back to the works of Charlie Hoare~\cite{hoareaxiom69}.
In this approach, the semantics or
mathematical meanings of a piece of code is defined through how
the piece of code modifies certain logic predicates on the variable(s)
of the code.

The basics of axiomatic semantics are demonstrated here using two examples.
In figure \ref{code01}, we have a piece of C code that computes the square of $x$
and assigns the answer to the variable $x$.
Notice in the comments or annotations that precede the code, we have two logic predicates $x<=0$ and $x>=0$, preceded by the symbols ``@ requires'' and ``@ ensures''.
They represent properties of $x$ that we claim to be true, respectively before and
after the execution of the line of code.
The keyword \emph{requires} denotes a \emph{pre-condition} and the keyword
\emph{ensures} keyword denotes a \emph{post-condition}.
The pre and post-conditions, together with the statement they surround, form a \emph{Hoare Triple}: 
they express a contract of sorts, namely, that for any execution of the program, regardless of what
has happened elsewhere, if the pre-condition is true before the statement is
executed, then the post-condition will be true after its execution. 
In this example, it is trivial to see that
if the variable $x$ is non-positive before the execution of
$x:=x*x$ then it will be non-negative afterwards.
However we stress here that any such contract inserted as annotations in the code
needs to be formally proven before it is said to be valid for the code.
\begin{figure}
\centering
\begin{lstlisting}
/*
  @ requires x<=0
  @ ensures x>=0
*/
{
         x=x*x;
}
\end{lstlisting}
\label{code01}
\caption{C code with ACSL}
\end{figure}

Consider the C implementation of a 1-dimensional linear state-space system in figure \ref{code02}.
The state-transition matrix $A$ is 0.98 and the input matrix $B$ is 0.02.
Unlike the previous example, this piece of code contains an infinite loop.
In the case of infinite loops, the difficulty lies in finding a property that will hold
both before, throughout and after the execution of the loop. Such properties are called
\emph{loop invariants}. They are closely related to invariant sets. We can express
such a property with a contract on the body of the loop where the pre- and post- conditions need to be identical.
It is often the case, for even trivial invariant, that verifying their correctness is a non-trivial task.
Comparatively speaking though,
verifying the correctness of a given invariant in an automatic fashion is still a more tractable task than
finding the invariants of a program automatically.
For even simple examples, it can be impossible for general automatic decision procedures to
compute invariants for the code without the application of domain specific knowledge.
In addition, there is also an assertion $input*input<1$ denoted by the keyword \emph{assumes}.
The difference between assertions and properties is that assertions are the assumptions that we make without
given any proofs for it.
In this case, we are going to assume the magnitude of the input variable is bounded by $1$.
Unlike the property $x*x<=1$, the assertion $input*input<1$ cannot be checked for its
correctness based only on the information available from the code.
For this example or any other linear state-space systems, we can apply domain specific knowledge, namely
Lyapunov-based theories, to compute an ellipsoid
invariant set $\mathcal{E}\l x, P \r=\lc x | x^{\m{T}} P x \leq 1 \rc$.
A collection of these type of quadratic stability results with an efficient computational solutions can be found
in~\cite{boydlmi94}.
For this example, $\mathcal{E} \l x,1\r$ forms a valid invariant set.
Through the credible autocoding framework,
this invariant property can be rapidly transformed into
contracts for the code, and thus,  in theory, makes the process of
automatic verification of the generated code more feasible.
\begin{figure}
\centering
\begin{lstlisting}
/*
  @ assumes input*input<1;
  @ requires x*x<=1
  @ ensures x*x<=1
*/
{
         while (1) {
               x=0.98*x+0.02*input;
         }
}
\end{lstlisting}
\label{code02}
\caption{C code with ACSL}
\end{figure}

The code annotations in the two examples are expressed in the ANSI C Specification Language (ACSL)
\footnote[1]{The prototype credible autocoder also produces annotations in ACSL}.
In the latter sections, the autocoded fault detection semantics are also expressed in ACSL. For more details,
interested readers can refer to ~\cite{baudinacsl08}.

\section{Fault detection problem formulation}

In this paper we focus on observer-based fault detection of dynamic systems. Such methods need the system to be modeled by differential equations. In this paper we design the fault detection observer for a three-degrees-of-freedom laboratory helicopter. The system is modeled by nonlinear equations.
Such a model can be linearized around the operating point of the system as follows
\begin{eqnarray}
\dot{x}(t)&=&A x(t)+B u(t)+ E f(t), \label{1}\\
y(t)&=&C x(t), \label{2}
\end{eqnarray}
where $x(t)\in \Re^6$ and $u(k) \in \Re^2$ are the state vector and the known input vector at time $t$, respectively.
Also, $y(t) \in \Re^{3}$ is the output vector .
$A$, $B$ and $C$ are state transition, input and output matrices, respectively:
\small
\begin{eqnarray*}
A\!\!\!\!\!\!&=&\!\!\!\!\!\!\begin{pmatrix}
0 & 0 & 0 & 1 & 0 & 0\\
0 & 0 & 0 & 0 & 1 & 0\\
0 & 0 & 0 & 0 & 0 & 1\\
0 & 0 & 0 & 0 & 0 & 0\\
0 & 0 & 0 & 0 & 0 & 0\\
0 & \frac{(2 m_f L_a-m_w L_m) g}{2 m_f L_a^2+2 m_f L_h^2+m_w L_m^2} & 0 & 0 & 0 & 0
\end{pmatrix}, \\
B\!\!\!\!\!\!&=&\!\!\!\!\!\!\begin{pmatrix}
0 & 0\\
0 & 0\\
0 & 0\\
\frac{L_a K_f}{(m_w L_w^2+2 m_f L_a^2)} & \frac{L_a K_f}{m_w Lw^2 +2 m_f L_a^2}\\
 \frac{K_f}{2 m_f L_f} &  \frac{-K_f}{2 m_f L_f}\\
0 & 0
\end{pmatrix}, \\
C\!\!\!\!\!\!&=&\!\!\!\!\!\!\begin{pmatrix}
1 & 0 & 0 & 0 & 0 & 0\\
0 & 1 & 0 & 0 & 0 & 0\\
0 & 0 & 1 & 0 & 0 & 0
\end{pmatrix}. \label{8}
\end{eqnarray*}
\normalsize
System parameters are given in \cite{Quanser}.
$f(t) \in \Re^{n_f}$ in \eqref{1} represents an additive fault to the system that should be detected. No prior knowledge on this input signal is available.
The value of $f(t)$ is zero for nominal (fault-free) system. The aim of the fault detection is to raise an alarm whenever this value differs significantly from zero (faulty system). For that purpose, an observers based fault detection method is used. Here, we do not develop new fault detection methods. Instead we focus on the software implementation of an observer-based fault detection method.

We consider an actuator degradation fault for this system. Such a fault changes the behavior and the steady state of the system, and can be modeled as additive fault. The effect of the degradation can be modeled by replacing $u(t)$ in \eqref{1} with $\bar{u}(t)$
where
\begin{eqnarray}
\bar{u}(t)= X u(t).
\end{eqnarray}
Hence, we obtain the fault matrix bellow, defined in \eqref{1}
\begin{eqnarray}
E=B(I-X).
\end{eqnarray}

\subsection{Output observer design for fault detection} \label{theory}

To explain the autocoding process, we select the simplest observer-based method \cite{ding2008model,isermann2005fault}.
The detector observes the system, received input and output data and compares the data with the
nominal response the system is suppose to have.

Consider the full-order state observer bellow
\begin{eqnarray}
\dot{\hat{x}}(t)&=&A \hat{x}(t)+B u(t)+ L (y(t)-C \hat{x}(t)), \label{3}\\
\hat{y}(t)&=&C \hat{x}(t). \label{3a}
\end{eqnarray}
Using this observer, we generate a residual signal, comparing the estimated output \eqref{3a} with the measured one
\begin{equation}
r(t)=y(t)-\hat{y}(t). \label{4}
\end{equation}
We compare the residual signal $r(t)$ against a predefined threshold. If the threshold is reached, a fault alarm is raised.
In order to explain how the method works and how the observer should be designed, we introduce estimation error $e(t)=x(t)-\hat{x}(t)$ and calculate the error dynamics
\begin{eqnarray}
\dot{e}(t)&=&(A-LC)e(t)+E f(t), \label{5}\\
r(t)&=&Ce(t). \label{5a}
\end{eqnarray}
From \eqref{5}--\eqref{5a}, $r(t)$ goes to zero if $f(t)$ is zero and
the observer matrix $L$ is chosen so that $A-LC$ is stable.
Note that $L$ is the only design parameter for this observer. In practice, usually we do not to raise a fault alarm if $f(t)$ is too small. Hence we suppose $\|f(t)\|>\sigma$ is a fault that must be detected. Consequently $\|r(t)\|>r_{th}$ raises a fault alarm, where $r_{th}$ is the threshold corresponding to $\sigma$.

\section{A formal method to verify fault detection} \label{veri}

The theory behind fault detection methods is presented in Section \ref{theory}. However, there always exist a semantic gap between theory and real implementation. The methods in Section \ref{theory} should be implemented in form of software, either in graphical environments of control design such as Simulink or as computer codes in computer languages such as Matlab or C. Due to the computation errors and replacing real numbers by floating numbers in digital computers, there exist a difference between implemented method in software form and the ideal results in theory. We aim at annotating the software so that an expert or a machine can track the operation of the software and verify that the design criteria are satisfied at software level. The idea developed in \cite{feron2010control} to formally document the stability of closed-loop systems is now extended to fault detection methods.

In order to certify the fault detection software, we need to certify particular properties of the observer
\begin{enumerate}
  \item Stability: the error dynamics is stable, i.e. $e(t)$ in \eqref{5} is around origin when system is in nominal case and stays bounded in faulty case.
  \item Fault detection: the residual $r(t)$ correctly detects the fault. In other words $r(t)$ dos not reach a predefined threshold if $f(t)$ is sufficiently small.
\end{enumerate}

To verify these properties, we use Lyapunov theory which was shown to be a very good mechanism to generate easy-to-use, formal code annotations (see \cite{feron2010control}). Note that checking the place of observer poles is not desired for software verification, because we need a mechanism to verify that each line of the code keeps invariant properties we prove in theory. Also, computer scientists and other engineers are familiar with invariant properties while understanding the connection between the place of system poles and convergent of control software needs a control theory background \cite{feron2010control}. On the other hand, formal verification methods in computer science work based on invariant sets.
For that purpose, we start from informal specifications in Section \ref{theory} and translate them into formal specification as follows. These  properties show how the software variables can change so that the pre-defined filter specifications are verified.

Suppose the system is in nominal mode. Considering a Lyapunov function $V(t)=e^T(t)Pe(t)$, where  $P$ is a positive definite matrix. We can show that the $e(t)$ remains in a predefined invariant ellipsoid
\begin{eqnarray}
\mathcal{E}_n=\{e(t)\in \Re^n | e^T(t)Pe(t) \leq \zeta \},\label{45a}
\end{eqnarray}
for all $t \in \Re$ if the observer is stable. Here, $\zeta\geq0$ is a scalar.

For the faulty mode we can introduce a similar ellipsoid around the new equilibrium point. However, we do not know the new equilibrium point, as the fault is supposed to be completely unknown. But in practice, $f(t)$ is bounded. Suppose that $\|f(t)\|<\sigma$. We introduce
\begin{eqnarray}
\mathcal{E}_f=\{e(t)\in \Re^n | e^T(t)Pe(t) \leq \bar{\zeta} \}.\label{46a}
\end{eqnarray}
In \eqref{46a} $\bar{\zeta}$ is
\begin{eqnarray}
\bar{\zeta}(t)&=& \max_e e^T(t)Pe(t) \nonumber\\
& & s.t. \;\;\; \dot{e}(t)=(A-LC)e(t)+E f(t) \nonumber\\
& & and \;\;\; \|f(t)\|<\sigma.\label{47a}
\end{eqnarray}
Hence, we have two ellipsoids to which the value of the Lyapunov function may belong. As far as $\forall t$, $V(t) \in \mathcal{E}_n$,  the system is in nominal mode and the observer is stable. On the other hand if $\forall t$, $V(t) \in \mathcal{E}_f$, the system is in faulty mode and the observer is stable. If $\exists t, V(t) \not\in \mathcal{E}_f$ the detector is unstable.

\begin{figure}[htbp]
\begin{center}
\includegraphics
[scale=0.25,angle=270]{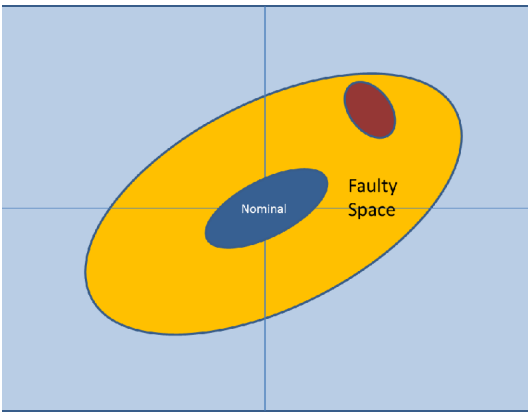}
\end{center}
\caption{Blue ellipsoid shows $\mathcal{E}_n$ and yellow ellipsoid demonstrates $\mathcal{E}_f$ in error space. Red ellipsoid is a fault scenario.}
\label{Lyapunov}
\end{figure}

Figure \ref{Lyapunov} shows  the nominal and faulty regions in the error space.

What remains is to relate the threshold on the residual signal to the value of $\zeta$. Here we do not explain the details. However, the following Lemma (or similar Lemmas) helps to calculate $\bar{\zeta}$.

\begin{lemma} \label{lm1}
given the system
\begin{eqnarray}
\dot{x}(t)=Ax(t) + E d(t),\;\; x(t)=0, \label{48}
\end{eqnarray}
for a given constant $\rho>0$ where $P$ is a positive definite symmetric matrix we have
\begin{eqnarray}
x^T(t)Px(t)<\rho \|d(t)\|, \label{49}
\end{eqnarray}
if there exist $Q>0$ so that
\begin{eqnarray}
\begin{pmatrix}
A^TQ+QA &Q E & P^{1/2}\\
E^TQ & -\rho I &0\\
P^{1/2} & 0 & -\rho I
\end{pmatrix}<0, \label{50}
\end{eqnarray}
\end{lemma}

\section{Autocoding of Fault Detection Semantics}
\begin{multicols}{2}
\begin{figure*}
\centering
\includegraphics[scale=0.7]{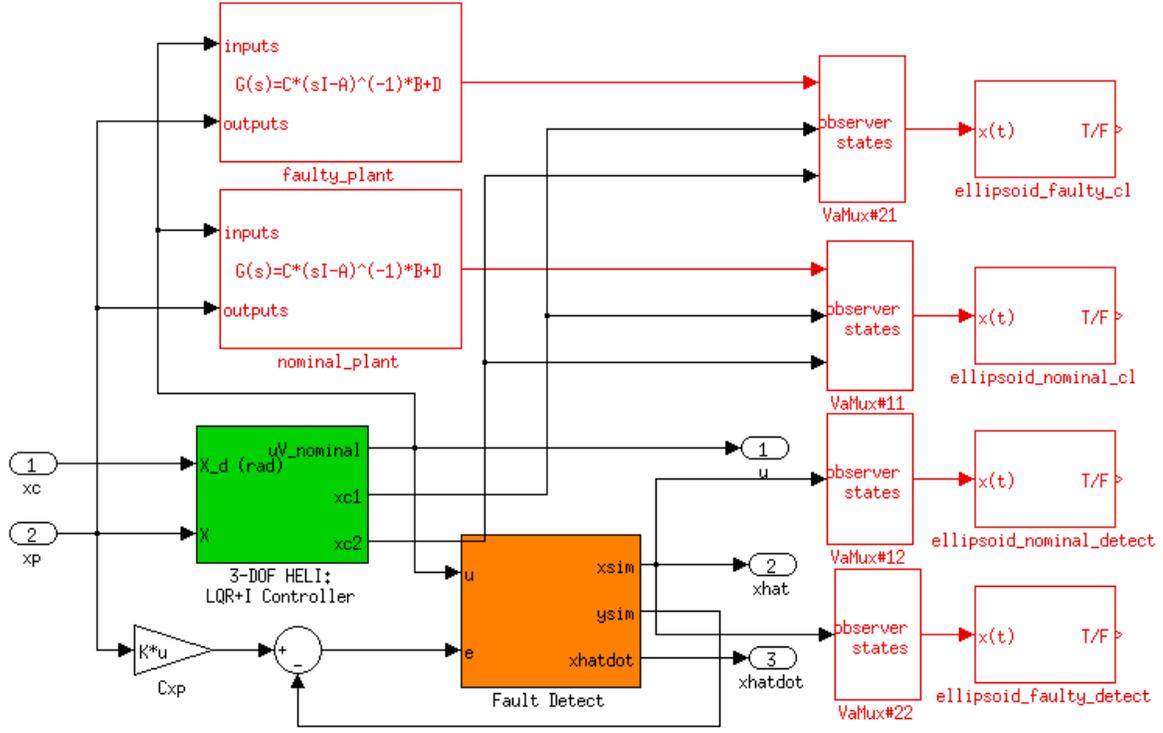}
\caption{Simulink Model Input For Credible Autocoding}
\label{complete}
\end{figure*}
\end{multicols}

In this section, we describe the autocoding of the fault detection semantics of a running example.
The running example is a fault detection system as specified in section \ref{theory} combined
with a LQR controller that has two integrators.
We first point out that on the abstraction level of a computer program, the notion of a
continuous-time differential equation like in (\ref{3a}) no longer applies.
The running example, including the plant, need to be in discrete-time.
In fact, for analysis purposes, the plant can be treated as another C program.
We have the following discrete-time linear state-space systems
\be
\ba{l}
\dps x_{c} (k+1) = A_{c} x_{c}(k)  + B_{c} y(k),  \cr
\dps u(k) = C_{c} x_{k} + D_{c} y(k),
\ea
\label{sys:01}
\ee
\be
\ba{l}
\dps \hat{x}(k+1)=\hat{A} \hat{x}(k) + B u + Ly,  \cr
\dps  r(k)=y(k)-\hat{y}(k),
\ea
\label{sys:02}
\ee
and
\be
\ba{l}
\dps x(k+1) = A x (k) + Bu(k),  \cr
\dps y(k) = Cx(k),
\ea
\label{sys:03}
\ee
representing the controller, the detector, and the plant respectively. After discretization, system matrices change. However, with an abuse of notation, we use the same symbols for the discrete-time model of the system and the observer in \eqref{sys:02}--\eqref{sys:03}.
We have also a discretized version of the error dynamics from (\ref{3a}),
\be
\dps e(k+1) = \hat{A} e(k) + E f(k)
\label{sys:04}
\ee
where $f(k)$ represent sampled fault signal and $\hat{A}=A-LC$.
The Simulink model of the controller and the fault detection system is displayed in Figure \ref{complete}.
Additionally, there are also fault detection semantics in the model.
They are expressed using the annotation blocks as displayed in red in Figure \ref{complete}.
The annotation blocks are converted into ACSL annotations for the output code.
The four Ellipsoid observer blocks represent four ellipsoid invariants.
Two are for the plant states $x$ and two are for the detector states $\hat{x}$.
The semantics of the plants (faulty and nominal)
are expressed using the Plant annotation blocks in Figure \ref{complete}.
We do not express the ellipsoid sets for the error states from (\ref{45a}) and (\ref{46a}) on the input Simulink model.
The reason for this is explained in section (\ref{why}).
However, we point out that the credible autocoding process will generate
two ellipsoid sets on the error states $e=x-\hat{x}$,
one for the nominal plant and the other for the faulty.
They are just expressed in the annotations of the generated code output.
Not shown in Figure \ref{complete} is the ellipsoid observer block expressing a bound on
the input signal $y_{c}$. This bound is an assumption that gets transformed into an ACSL assertion.

\subsection{Ellipsoid sets on the error states $e$}
\label{why}

In this section, we discuss the reason behind not choosing the
the two ellipsoid sets from (\ref{45a}) and (\ref{46a}) to be expressed in the
the input Simulink model, directly (instead we approximate it from the bounds on system and observer states in the code).
Consider the following two dynamics
\be
\dps \hat{x}(k+1) = \hat{A} \hat{x} (k) + B u(k) + LCx(k),
\label{sys:error1}
\ee
and
\be
\dps \hat{x}(k+1) = A x(k) + Bu(k) + LC e(k).
\label{sys:error2}
\ee
Note that (\ref{sys:error1}), (\ref{sys:error2}), and (\ref{sys:02}) are equivalent.
In the credible autocoding process, any ellipsoid invariant
that is inserted into the code from the input model is propagated
forward through the code using ellipsoid calculus~\cite{ellipsoid97}.
Methods such as computing the affine transformation of
an ellipsoid are very useful here, since semantically speaking,
most of the generated code is consisted of assigning affine expressions to variables.
For example, if an ellipsoid set
$\mathcal{E} \l x, P \r$ is the pre-condition, and the ensuing
block of code is semantically $x:=A x$,
then the autocoder generates the post-condition
$\mathcal{E} \l x, \l A P^{-1}A^{\m{T}} \r^{-1} \r$.
On the actual C code, the propagation steps are much smaller
so there are a sequence of intermediate ellipsoid invariants between
$\mathcal{E} \l x, P \r$ and $\mathcal{E} \l x, \l A P^{-1}A^{\m{T}} \r^{-1} \r$.
Let $\tilde{x}=\dps \lb \ba{c} x \cr x_{c} \ea \rb$ be the closed-loop system states,
and assume that closed-loop stability
analysis yields an ellipsoid invariant set $\mathcal{E} \l \tilde{x}, P_{0} \r$.
Given $\mathcal{E} \l \tilde{x}, P_{\tilde{x}} \r$, the prototype
autocoder can generate $P_{u}$, $P_{x}$ such that
$\mathcal{E} \l u, P_{u} \r$ and $\mathcal{E} \l x, P_{x}\r$.
Given $\mathcal{E} \l u , P_{u} \r$, $\mathcal{E} \l x, P_{x}\r$, and the dynamics
in (\ref{sys:error1}), one can compute an ellipsoid invariant
$\mathcal{E} \l \hat{x}, P_{\hat{x}} \r$ for the detector states $\hat{x}$ by solving a linear matrix inequality.
Solving the LMI also yields the relaxation multipliers $\alpha>0$ and $\gamma>0$
for the quadratic inequalities in $\mathcal{E} \l u , P_{u} \r$ and $\mathcal{E} \l x, P_{x}\r $.
These multipliers are used to generate an ellipsoid invariant on $e$ in the following way.
Given the ellipsoid invariants $\mathcal{E} \l \hat{x}, P_{\hat{x}} \r $, $\mathcal{E} \l x, P_{x}\r$, and
the error states $e= \lb \ba{cc} I & -I \ea \rb \dps \lb \ba{c} x \cr \hat{x} \ea \rb$, a correct ellipsoid invariant on $e$ is $\mathcal{E} \l e, P_{e} \r$, 
where
\be
\dps P_{e} = \l \lb \ba{cc} I & -I \ea \rb  P_{x,\hat{x}}^{-1} \lb \ba{cc} I & -I \ea \rb^{\m{T}} \r^{-1}
\label{sproc}
\ee with $P_{x,\hat{x}} = \lb \ba{cc} \gamma  P_{x} & 0 \cr 0 & \l 1-\alpha-\gamma \r
P_{\hat{x}} \ea \rb$.
By choosing to express
the ellipsoid invariants $\mathcal{E} \l \hat{x}, P_{\hat{x}} \r $, $\mathcal{E} \l \tilde{x}, P_{\tilde{x}}\r$ on the input Simulink model, the autocoder can automatically produce ellipsoid invariants on the error states $e$.

Alternatively, if we choose to instead to express
the ellipsoids on $e$ 
and $\mathcal{E} \l \tilde{x}, P_{\tilde{x}}\r$ on the Simulink model,
then the autocoder encounters the problem of
computing an ellipsoid invariant for $\hat{x}$ given the dynamics in (\ref{sys:error2}),
with the assumptions of $\mathcal{E} \l \tilde{x}, P_{\tilde{x}} \r$ and $ \mathcal{E} \l e, P_{e} \r$.
This is infeasible since $A$ is not stable in this example.
Without an invariant set for $\hat{x}$, there are no safety bounds on the
program variables that correspond to $\hat{x}$.

In the first option, the annotations generated by the credible autocoding process can
guarantee both the safety property (the ellipsoid bounds on 
the variables correspond to $x_{c}$ and $\hat{x}$) and the liveness property of the fault
detection system i.e. the two ellipsoids on the error states. 
In the second option, they can only guarantee the latter.

\subsection{Ellipsoid sets in the Simulink model}

To generate the ellipsoid invariants for the credible autocoding,
we have
\be
\dps \hat{x}(k+1) = \hat{A} \hat{x}(k) + \hat{B} \hat{u}(k)
\label{fddyn}
\ee
with $\tilde{B}=LC$, $\hat{A} = A - LC$, $\dps \hat{u} =\lb \ba{c} u \cr x \ea \rb$,
and $\hat{B} = \lb \ba{cc} B & \tilde{B} \ea \rb$.
Given that the closed-loop ellipsoid set $\mathcal{E} \l \tilde{x}, P_{\tilde{x}}\r $ implies $\mathcal{E} \l u, P_{u} \r$
and $\mathcal{E}  \l x ,P_{x}\r$ for some matrices $P_{u}$, $P_{x}$ by the affine transformation of ellipsoid set.
With $\mathcal{E} \l u, P_{u} \r$,
$\mathcal{E}  \l x ,P_{x}\r$, and the detector dynamics in (\ref{sys:error1}), 
we have the following results for computing an ellipsoid invariant on $\hat{x}$.
\begin{lemma}
Let $\hat{u} = \lb \ba{c} u \cr x \ea \rb $ and assume that $\hat{u}$ belongs to the
set $\lc \hat{u} | \hat{u}^{\m{T}} P_{1} \hat{u} \leq 1 \rc$.
If there exist a symmetric positive-definite matrix $P$ and a positive scalar $\alpha$
that satisfies the following linear
matrix inequality
\be
\dps \lb \ba{cc}
\hat{A}^{\m{T}} P \hat{A} - P + \alpha P  &  \hat{A}^{\m{T}} P \hat{B} \cr
\hat{B}^{\m{T}} P \hat{A} &  \hat{B}^{\m{T}} P \hat{B} - \alpha P_{1}
\ea \rb \prec 0
\label{lmi02}
\ee
then the set $\lc \hat{x} | \hat{x}^{\m{T}} P \hat{x} \leq 1 \rc$ is invariant with respect to (\ref{fddyn}).
\label{lemma02}
\end{lemma}

First we manually compute the invariant sets for the closed-loop system. Once for the faulty plant
and one more for the nominal plant using similar techniques as lemma \ref{lm1}.
For the closed-loop analysis, we assume the command input $y_{c}$ is bounded.
From the obtained closed-loop invariant sets, the autocoder can generate two ellipsoid invariants on $\hat{u}$.
With $\mathcal{E} \l \hat{u}_{i}, P_{\hat{u}_i}, i=N,F\r$, and $N,F$ denotes respectively nominal or faulty.
Now we apply lemma \ref{lemma02} twice to obtain the two ellipsoid sets
$\mathcal{E} \l \hat{x}_{i}, P_{\hat{x}_{i}} \r, i=N,F$ on $\hat{x}$.
As discussed before, we insert the obtained ellipsoid invariants
$\mathcal{E} \l \tilde{x}_{n}, P_{\tilde{x}_i} \r,i=N,F$ and
$\mathcal{E} \l \hat{x}_{i}, P_{\hat{x}_i} \r,i=N,F$ on the detector states $\hat{x}$
into the Simulink model. The ellipsoid invariants on $e$ are automatically 
computed by the credible autocoder using (\ref{sproc}), thus does not need to be 
expressed on the Simulink model. 

\subsection{Prototype Refinements and the Annotations}
For the automatic transformation of the semantics of the fault detection and controller system
in Figure \ref{complete} into useful ACSL annotations,
we have further refined the prototype autocoder to be able handle the following issues:
\begin{enumerate}
\item Generate different sets of closed-loop semantics based on different assumptions of the plant.
\item Formally expressing the faults to be able to reason about them in the invariant propagation process.
\end{enumerate}
The main change made to the prototype
is a new capability to generate multiple different sets of
closed-loop semantics based on the assumptions of the different plant semantics.
For example, in the generated
ACSL annotated code in listing \ref{acsl01}, there are two
ellipsoid sets parameterized by the
the ACSL matrix variables \emph{QMat\_1} and \emph{QMat\_2}.
They express the closed-loop ellipsoid invariant
sets $\mathcal{E} \l \hat{x}_i, P_{\hat{x}_i}\r,i=N,F$.
The matrix variables are assigned the correct values using the
ACSL functions \emph{mat\_of\_$n$x$n$\_scalar}, which takes in $n^2$
number of real-valued arguments and returns an array of size $n \times n$.
For brevity's sake, the input arguments to
the ACSL functions in listing \ref{acsl01} are truncated.
The ellipsoid sets are grouped into two different set of semantics using the ACSL keyword
\emph{behavior}. One set of semantics assumes a nominal plant and the other assumes the faulty.
Each set of semantics are linked to their respective plant models by the behavior name.
The pre-conditions displayed in listing \ref{acsl01}
are ellipsoid invariant sets on the observer states $\hat{x}$ defined by
the ACSL variables \emph{QMat\_3} and \emph{QMat\_4}.
The post-conditions are generated using the invariant propagation process
as described in section \ref{why}.
The annotation statement \emph{PROOF\_TACTIC} is
a non-ACSL element that the prototype autocoder generates to assist the
automatic verification of the invariants. For example, to formally prove that the post-conditions
in listing \ref{acsl01} is true given the pre-conditions, the automatic analyzer
knows from the \emph{PROOF\_TACTIC} statement to apply the affine transformation strategy.
Subsection \ref{autoverify} has more details on the automatic verification of the annotations.
\begin{lstlisting}[caption={ACSL Expressing Multiple Sets of Closed-loop Semantics},label={acsl01}]
/*@
	logic matrix QMat_1=mat_of_8x8_scalar(...);
*/
/*@
	logic matrix QMat_2=mat_of_8x8_scalar(...);

*/
/*@
	logic matrix QMat_3 = mat_of_6x6_scalar(...)
*/
/*@
        logic matrix QMat_4 = mat_of_6x6_scalar(...)
*/

/*@
	behavior nominal_ellipsoid:
	requires in_ellipsoidQ(QMat_3,
	vect_of_6_scalar(observer_states[0],
	observer_states[1],observer_states[2],
	observer_states[3],observer_states[4],
	observer_states[5]));
	ensures in_ellipsoidQ(QMat_41,
	vect_of_12_scalar(observer_states[0]...,
	_io_->xhat[0]...));
      	@ PROOF_TACTIC (use_strategy (AffineEllipsoid));
*/
/*@
        behavior faulty_ellipsoid:
	requires in_ellipsoidQ(QMat_4,
	vect_of_6_scalar(observer_states[0],
	observer_states[1],observer_states[2],
	observer_states[3],observer_states[4],
	observer_states[5]));
	ensures in_ellipsoidQ(QMat_42,
	vect_of_12_scalar(observer_states[0]..._io_->xhat[0]..));
      	@ PROOF_TACTIC (use_strategy (AffineEllipsoid));
*/
{
    for (i1 = 0; i1 < 6; i1++) {
        _io_->xhat[i1] = observer_states[i1];
    }
}
\end{lstlisting}

The semantics of the plant models are expressed using the C functions \emph{faulty\_plant} and \emph{nominal\_plant}
in the ghost code statements denoted by the \emph{ghost} keyword.
Ghost code statements are ACSL statements that are similar to the actual C code in every aspect
except they are not executed and they are restricted from changing the state of any variables in the code.
The semantics of the plant model
are connected with their respective set of ellipsoid invariants
through the usage of assertions. For example, in listing \ref{acsl02}, we have
the plant states \emph{faulty\_state} and \emph{nominal\_state}, which are declared in the ghost code statements.
They are linked to the same variable \emph{\_io\_-$>$xp} from the code
in the two ACSL behaviors using the equal relation symbol $==$.
\begin{lstlisting}[caption={ACSL Expressing the Semantics of the Plants with C code},label={acsl02}]
/*@
	ghost double faulty_state[6];
	ghost double nominal_state[6];
*/
/*@
        behavior nominal_ellipsoid:
        assumes _io_->xp==nominal_state
        requires in_ellipsoidQ(QMat_1,
	vect_of_8_scalar(..._));
	ensures in_ellipsoidQ(QMat_7,
	vect_of_14_scalar(...));
       @ PROOF_TACTIC (use_strategy (AffineEllipsoid));
*/
/*@
        behavior faulty_ellipsoid:
	assumes _io_->xp==faulty_state
	requires in_ellipsoidQ(QMat_2,
        vect_of_8_scalar(...));
        ensures in_ellipsoidQ(QMat_8,
        vect_of_14_scalar(...));
        @ PROOF_TACTIC (use_strategy (AffineEllipsoid));
*/

{
    for (i1 = 0; i1 < 6; i1++) {
	    xp[i1] = _io_->xp[i1];
    }
}
.
.
.
/*@
	ghost faulty_plant(_io_->u,faulty_plant_state);
	ghost nominal_plant(_io_->u,faulty_plant_state);
*/
\end{lstlisting}

Finally the key property of fault detection is generated by a special annotation
block that indicates to the autocoder, the specific ghost variable that we want to express ellipsoid invariants
for. In this case, the ghost variable of interest is the one that corresponds to the
error states $e=x-\hat{x}$. The variable is a ghost variable since the error states
$e$ do not explicitly correspond to any variables generated in the code. The autocoder inserts the definition of the
error states in the form of another ghost code statement as shown in listing \ref{acsl03}.
As in the previous two code snippets, there are two different set of
ellipsoid invariants on the variables error\_states[i1], which is
based on initial plant model assumptions of either faulty or nominal.
\begin{lstlisting}[caption={ACSL Expressing Invariant Sets on the Error States},label={acsl03}]

/*@	ghost for (i1=0; i1<6; i1++) {
		error_states[i1]=x[i1]-observer_states[i1];
	}
*/
/*@
	 behavior nominal_ellipsoid:
	 requires in_ellipsoidQ(QMat_30,
	 vect_of_12_scalar(x[0],...,observer_states[0],...));
	 ensures in_ellipsoidQ(QMat_31,
	 vect_of_6_scalar(error_states[0],...));
	@ PROOF_TACTIC (use_strategy (AffineEllipsoid));

*/
/*@
        behavior faulty_ellipsoid:
	requires in_ellipsoidQ(QMat_32,
	vect_of_12_scalar(x[0],...,observer_states[0],...));
	ensures in_ellipsoidQ(QMat_33,
	vect_of_6_scalar(...));
	@ PROOF_TACTIC (use_strategy (AffineEllipsoid));

*/

{
    for (i1 = 0; i1 < 6; i1++) {
        observer_states[i1] = _state_->observer_states_memory[i1];
    }

    for (i1 = 0; i1 < 6; i1++) {
        x[i1] = _io_->x[i1];
    }
}
\end{lstlisting}

\subsection{Automatic Verification} \label{autoverify}
In ~\cite{wjf13arxiv}, a \emph{backend} to the prototype
autocoder is developed. It takes as input the annotated C code generated by the
autocoder and outputs a certificate of validity in the form of proofs of correctness for the annotations.
The Frama-C/WP platform~\cite{framac}~\cite{wp} reads the annotated code and generates logic properties which truth is equivalent
to the correctness of the annotations. The Why3 tool~\cite{boogie11why3} converts these properties into a format readable by
the interactive theorem prover PVS~\cite{PVS-CADE92}. The annotation statement \emph{PROOF\_TACTIC} provides
our framework with the necessary information to generate the proof to these properties and check them in PVS.

While the prototype autocoder needed to be extended to handle fault detection
software, the nature of the annotations on the C code and the type of
logic reasoning used to prove them remained the same.
As such, this backend is readily available to verify
the correctness of the generated annotations. However this verification
is currently done under the assumption that computations occurring in
the program return the exact real value, and not the floating point approximation,
of their result. This extension is left for future research.
\section{Conclusion}
In this paper, we have presented a framework that can rapidly generate fault detection
code with a formal assurance of high-level fault detection semantics such as stability and
correct fault detection.
The properties are formally expressed using ellipsoid invariants.
The framework, dubbed credible autocoding, can generate the fault detection code as well as the invariant
properties for the code.
Moreover, the generated invariant properties can be
be verified in using semi-automatic theorem prover.
We have demonstrated that the credible autocoding prototype that was previously applied
to control systems can be extended to fault detection systems with some additions.
We applied the prototype tool to an example of observer-based fault detection
system running with a LQR controller of a $3$ degrees-of-freedom helicopter.
The prototype was able to autocode the fault detection semantics successfully.
In this paper we only consider a simple output observer for fault detection to demonstrate the autocoding steps. However, the idea can be extended 
to more complicated fault detection methods.

\bibliographystyle{IEEEtran}
\bibliography{IEEEabrv,complete,credible,biblio}

\end{document}